\documentclass[reprint,aps,prl,twocolumn,floatfix]{revtex4-2}

\usepackage{graphicx}
\usepackage{siunitx}
\usepackage{gitinfo2}
\usepackage{amsmath}
\usepackage{soul}
\usepackage{hyperref}

\usepackage[caption=false]{subfig}
\newlength{\onecolfig}
\newlength{\twocolfig}
\newcommand{\alicefidelity}{97.90(12)\%{}}
\newcommand{\bobfidelity}{97.70(12)\%{}}
\newcommand{\abfidelity}{94.0(5)\%{}}

\newcommand{\absuccessp}{\num{2.18e-4}}

\newcommand{\alicerate}{\SI{4.0e3}{\per\second}}
\newcommand{\bobrate}{\SI{5.7e3}{\per\second}}
\newcommand{\abrate}{\SI{182}{\per\second}}

\newcommand{\ion}[2]{\mbox{$^{#2}$#1$^+$}}

\newcommand{\Sr}[1]{\ion{Sr}{#1}}

\newcommand{\lev}[2]{{#1}_{#2}}
\newcommand{\fslev}[3]{{#1}_{#2},\,m\!=\!{#3}}  %

\newcommand{\ish}{\mbox{$\sim$}\,}

\newcommand{\ket}[1]{\left| #1 \right>}
\newcommand{\ketfixed}[1]{\big| #1 \big>}  %

\newcommand{\up}{\ket{\uparrow}}
\newcommand{\down}{\ket{\downarrow}}
\newcommand{\dstate}{\ket{D}}

\begin{document}

\title{High-rate, high-fidelity entanglement of qubits across an elementary quantum network}

\author{L.~J.~Stephenson}
\thanks{These two authors contributed equally}
\author{D.~P.~Nadlinger}
\thanks{These two authors contributed equally}
\author{B.~C.~Nichol}
\author{S.~An}
\author{P.~Drmota}
\author{T.~G.~Ballance}
\thanks{Present address: ColdQuanta UK Ltd, Oxford}
\author{K.~Thirumalai}
\author{J.~F.~Goodwin}
\author{D.~M.~Lucas}
\author{C.~J.~Ballance}
\email{chris.ballance@physics.ox.ac.uk}
\affiliation{Department of Physics, University of Oxford, Clarendon Laboratory, Parks Road, Oxford OX1 3PU, U.K.}

\date{May 12, 2020}

\begin{abstract}
We demonstrate remote entanglement of trapped-ion qubits via a quantum-optical fibre link with fidelity and rate approaching those of local operations.
Two \Sr{88} qubits are entangled via the polarisation degree of freedom of two spontaneously emitted \SI{422}{\nano\metre} photons which are coupled by high-numerical-aperture lenses into single-mode optical fibres and interfere on a beam splitter.
A novel geometry allows high-efficiency photon collection while maintaining unit fidelity for ion-photon entanglement.
We generate heralded Bell pairs with fidelity 94\%\ at an average rate \abrate{} (success probability \absuccessp{}).
\end{abstract}

\maketitle

The ability to form logical connections between all quantum bits (qubits) of a quantum processor is a prerequisite for building a fault-tolerant universal device~\cite{DiVincenzo2000}.
Trapped atomic ions have been identified as an excellent candidate qubit technology because they allow the implementation of single-qubit operations~\cite{Harty2014,Ballance2016,Gaebler2016}, two-qubit phonon-mediated gates~\cite{Ballance2016,Gaebler2016} and quantum memories~\cite{Wang2017,Sepiol2019}, all with high fidelity.
However, the number of ions that can be reliably interfaced in a single trap is limited by the motional mode density, necessitating architectures with multiple trap zones each hosting comparatively few ions.
Trap zones can be interfaced by physically shuttling qubits across centimetre-scale distances using electric fields~\cite{Kielpinski2002}, or by using photons to distribute entanglement over larger distances~\cite{Monroe2013}.
Photonic entanglement could also increase the connectivity of trapped-ion qubits via dynamically switchable fibre links~\cite{Kim2011a}, or allow the interfacing of different qubit platforms~\cite{Meyer2015}.
It also enables other quantum networking applications such as quantum key distribution, teleportation of quantum states, and blind quantum computing~\cite{Kimble2008,Broadbent2009}.
For ions, the entanglement rate is limited fundamentally only by the photon scattering rate (\ish\SI{100}{\mega\hertz}), exceeding local multiqubit operation rates (motional gates~\cite{Schafer2018a} and shuttling~\cite{Walther2012,Bowler2012}) at typical secular trap frequencies (\ish\SI{1}{\mega\hertz}).
In practice, photonic entanglement rates have been far lower than this, limited principally by low photon collection efficiencies~\cite{Monroe2014}; the highest previously reported rate for ions was \SI{4.5}{\per\second}, with 78\%\ fidelity~\cite{Hucul2015}.
Faster rates have been achieved with nitrogen-vacancy centres (\SI{39}{\hertz}) and quantum dots (\SI{7.3}{\kilo\hertz}), with fidelity $\approx 60\%$ \cite{Humphreys2018,Stockill2017}. Heralded entanglement of remote qubits with fidelity above 90\%\ has not previously been reported for any physical systems at rates above a few millihertz~\cite{Lettner2011,Ritter2012,Hensen2015}.

In this Letter, we report the generation of entanglement between two qubits in separate ion traps at rates and fidelities approaching those of typical local (intratrap) operations, by swapping entanglement between photons emitted by the ions onto the ions themselves~\cite{Zukowski1993}.
At these higher rates and fidelities, distillation procedures based on photonic entanglement~\cite{Nigmatullin2016} start to become a viable method for creating high quality entanglement across a scalable trapped-ion quantum computer.

A novel excitation scheme using \Sr{88} ions with photon collection perpendicular to the static applied magnetic field allows an increased rate over previous experiments~\cite{Hucul2015}, with polarisation mixing maximally suppressed by coupling into a single mode optical fibre.
In contrast to previous schemes using \ion{Yb}{171}, the collection geometry does not impede the use of beams parallel to the applied magnetic field.
This allows standard $\sigma$-polarised optical pumping to be employed, thus permitting a wider choice of ion species and the straightforward initialisation of multiple ion species in a single trap.

\begin{figure}
  \centering
  {\footnotesize(a)\hfill}

  \includegraphics[width=\onecolfig, trim={0 2mm 0 5mm}]{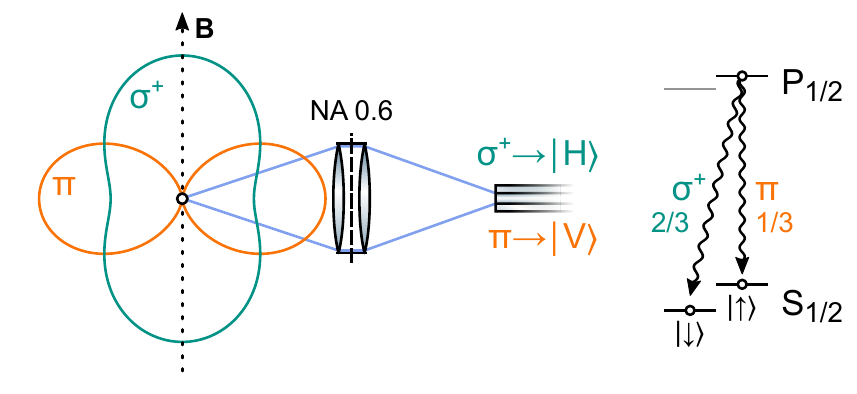}
  {\footnotesize(b)\hfill}

  \includegraphics{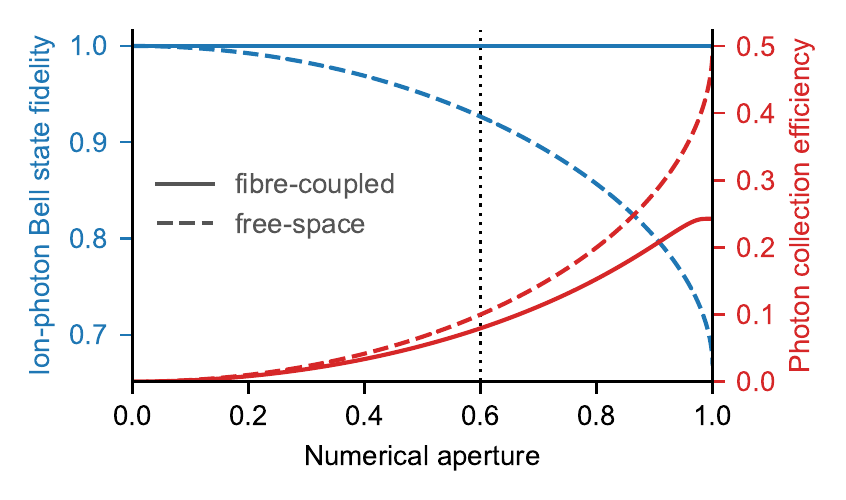}
  \caption{%
    (a) Intensity distribution of the light field for $\sigma^+$ and $\pi$ decay channels, relative to the quantisation axis set by the static magnetic field $\mathbf{B}$, and the branching fractions for each decay due to atomic selection rules.
    At the fibre input face, photons from $\sigma^+$ ($\pi$) decay map to the $\rm{H}$ ($\rm{V}$) fibre polarisation mode.
    (b) Maximum fraction of photons emitted by the ion that can be collected (red) and theoretical ion-photon entanglement fidelity (blue) versus collection optic numerical aperture.
    For free-space collection (dashed lines) polarisation mixing leads to a loss in fidelity with increasing numerical aperture.
    To model the fibre-coupling (solid lines), we calculate the overlap of the light field with a Gaussian mode on the dashed plane in (a): a smaller fraction of the emission is collected, but the polarisation mixing is completely suppressed.
    The vertical dashed line shows the NA \num{0.6} used in this Letter, where the fibre collection efficiency is 80\%\ of that in free space.
  }
  \label{F:Collection}
\end{figure}

We collect photons from the spontaneous decay of the excited electronic state $\ketfixed{\fslev{5p\,P}{1/2}{+1/2}}$ of \Sr{88}, as shown in Fig.~\ref{F:Collection}(a).
Decays to the two states of the ground level $\lev{5s\,S}{1/2}$ are associated with $\pi$ and $\sigma^+$ polarised photons, forming an entangled ion-photon state given by
\begin{equation*}
\ket{\psi} = \sqrt{\tfrac{2}{3}}\: \down\ket{\sigma^+} + \sqrt{\tfrac{1}{3}}\: \up\ket{\pi}\,,
\end{equation*}
where the weightings are due to the Clebsch-Gordan coefficients for each decay path, and the ion qubit states are labeled with $\down$ and $\up$.
Perpendicular to the magnetic field axis, the emitted field from the $\pi$ decay has twice the intensity, and so for photons on the collection axis the ion-photon state is
\begin{equation}
  \ket{\psi} = \tfrac{1}{\sqrt{2}}\, \big( \down\ket{H} + \up\ket{V} \!\big) \,,
  \label{E:ionphoton}
\end{equation}
where $\sigma^+$ and $\pi$ have been relabeled $H$ and $V$ to emphasise that the two photon polarisations are both linear and orthogonal; note that this is a maximally entangled Bell state.

The nonorthogonality of the $\sigma^+$ and $\pi$ emissions away from the collection axis would normally reduce the fidelity of the ion-photon entanglement at the high numerical apertures needed to maximise the photon collection efficiency~\cite{Blinov2004}.
However, with the chosen collection geometry, coupling into a single mode optical fibre rejects the nonorthogonal component of the $\sigma^+$ emission, reducing the maximum possible collection efficiency but maintaining unit ion-photon Bell state fidelity independent of collection aperture [see Fig.~\ref{F:Collection}(b)].
In contrast to other schemes, no photons of comparable wavelength are produced from undesired decay channels.
This eliminates the need to filter out such photons~\cite{Moehring2007}, enabling higher rates to be achieved with our collection geometry and excitation scheme.

By collecting two such photons entangled with separated ions and erasing the which-path information from the photons, a projective measurement of the two-photon state in the Bell basis will herald the projection of the two ions into a corresponding Bell state~\cite{Simon2003}.

In our experiment, \Sr{88} ions are trapped in two identical, high-optical-access, microfabricated surface traps~\footnote{Sandia National Laboratories HOA2.} in two vacuum systems, designated ``Alice'' and ``Bob'', separated by \SI{2}{\metre}.
In each system, a high-numerical-aperture (NA \num{0.6}) lens, aligned perpendicular to the applied magnetic field of \SI{0.56}{\milli\tesla}, couples single photons from the ion into an antireflection (AR) coated single-mode optical fibre.
Non-polarisation-maintaining (non-PM) fibres are used so as to introduce minimal differential phase between $H$ and $V$ photons (PM fibres introduce a large, temperature-sensitive, differential phase which would be difficult to control).
A second objective (NA \num{0.3}) images the ion through a slot in the trap onto a photomultiplier tube for  fluorescence detection.

\begin{figure}
  \includegraphics[width=\onecolfig]{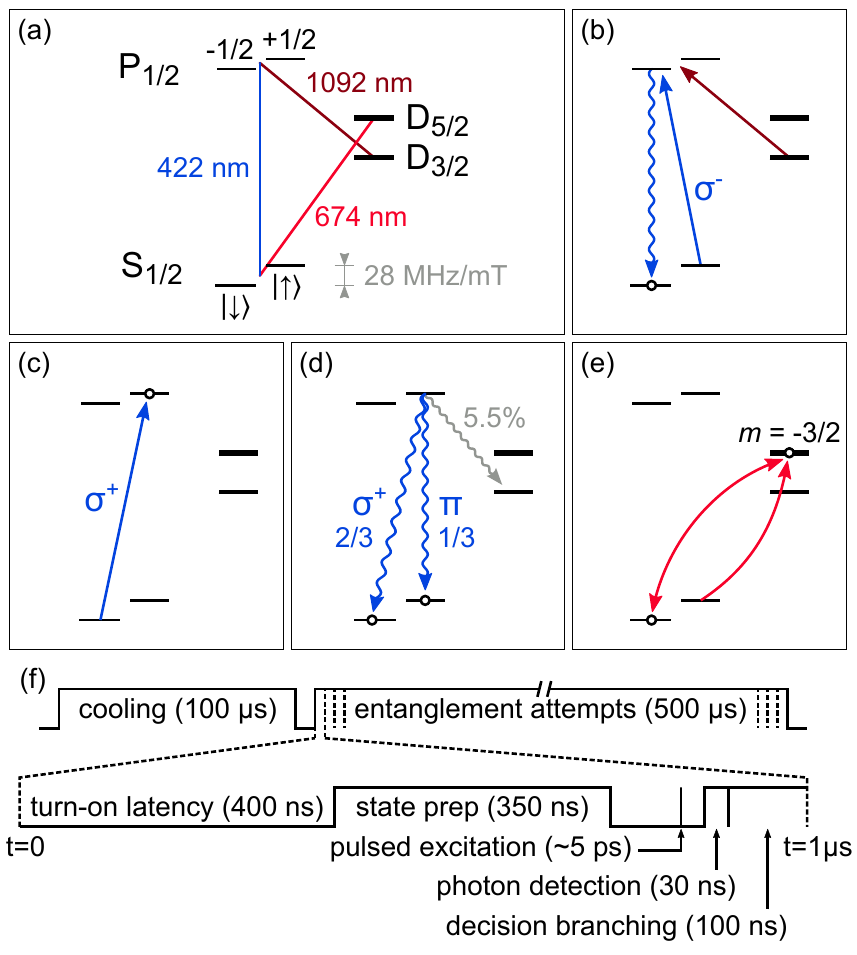}
  \caption{%
    (a) \Sr{88} level diagram (not to scale).
    (b) The initial state preparation consists of optical pumping on the \SI{422}{\nano\metre} transition, with a repumper at \SI{1092}{\nano\metre} to clear the $\lev{D}{3/2}$ level.
    (c) A single \ish\SI{5}{\pico\second} pulse from a frequency-doubled mode-locked Ti:sapphire laser coherently transfers the population to $\fslev{P}{1/2}{+1/2}$ with $\approx 97\%$ probability.
    (d) The ion decays to a superposition of $\down$ and $\up$, emitting a photon whose polarisation state is entangled with the state of the ion.
    Decays to the $\lev{D}{3/2}$ manifold occur with probability \SI{5.5}{\percent}, but as the \SI{1092}{\nano\metre} photons are not transmitted by the fibre, the only effect is to lower the overall rate.
    (e) Coherent manipulations are performed on the \SI{674}{\nano\metre} transition to $\dstate$ in order to analyse the final ion qubit state.
    (f) Experimental sequence: the ions are Doppler cooled for \SI{100}{\micro\second} before the attempt loop (lasting up to \SI{500}{\micro\second}) begins.
    The enlarged view shows a single attempt, with $\approx\SI{400}{\nano\second}$ of latency between state preparation turn-on signal (at $t=0$) and light arriving at the ion.
    State preparation $(\approx\SI{350}{\nano\second})$ is followed by a \SI{100}{\nano\second} delay to ensure that the beams are fully extinguished before the pulsed excitation.
    The \SI{30}{\nano\second} photon detection window begins \SI{30}{\nano\second} after the excitation pulse to allow for detector latency.
    A further \SI{100}{\nano\second} is required to decide whether to branch out of the attempt loop, in the event that a herald pattern is detected.
  }
  \label{F:SrLevels}
\end{figure}

The relevant electronic structure of \Sr{88} is shown in Fig.~\ref{F:SrLevels}.
Ions are Doppler cooled with lasers at \SI{422}{\nano\metre} and \SI{1092}{\nano\metre}.
The Zeeman structure of the ground level is used to encode the ``Zeeman'' qubit: $\ket{\fslev{S}{1/2}{-1/2}} =: \down$ and $\ket{\fslev{S}{1/2}{+1/2}} =: \up$.
We also define an ``optical'' qubit between the metastable level $\ket{\fslev{4d\,D}{5/2}{-3/2}} =: \dstate$ and $\down$, and use a narrow linewidth laser at \SI{674}{\nano\metre} to coherently transfer population between either of the Zeeman qubit states and $\dstate$, for ion state tomography.
As $\dstate$ is outside the Doppler cooling cycle, it can also be used to shelve population from $\up$ to measure the ground state qubit by state-dependent fluorescence detection~\cite{Myerson2008}.

The experimental sequence for generating entangled photons is shown in Fig.~\ref{F:SrLevels}.
An optimised attempt section at rate \SI{1}{\mega\hertz}, lasting at most \SI{500}{\micro\second}, is interleaved with \SI{100}{\micro\second} of Doppler cooling, until detection of an appropriate two-photon coincidence heralds the creation of ion-ion entanglement.
(In single-ion/single-photon experiments, a single click of a chosen detector instead breaks this attempt loop and triggers the start of the analysis sequence.)
The sequence is controlled by an FPGA~\footnote{ARTIQ Sinara hardware, \url{https://m-labs.hk/experiment-control/sinara-core/}}, incorporating the custom-optimised, precompiled section with decision branching in hardware, and just-in-time compiled sequences for qubit manipulations.

\begin{figure}
  \includegraphics[width=\onecolfig]{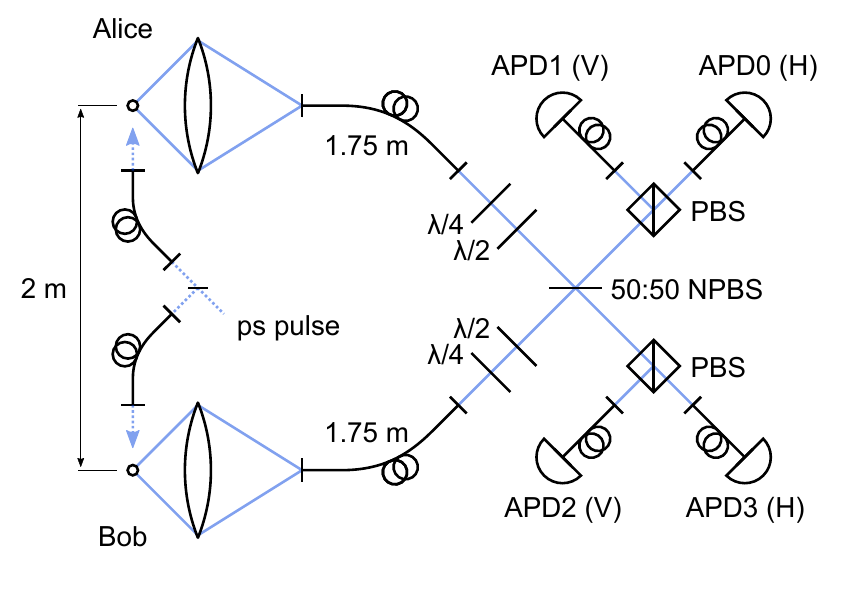}
  \caption{%
    Overview of apparatus.
    Single pulses from a frequency-doubled mode-locked laser are split to simultaneously excite ions in Alice and Bob.
    The spontaneously emitted photons are collected into single-mode non-PM fibres, with wave plates directly after the fibres to rotate the polarisation.
    The photons are then overlapped on a 50:50 NPBS, and detected on single-mode-fibre-coupled APDs following PBSs.
    To minimise the polarisation dependence of the NPBS, we use a small angle of incidence ($\approx\ang{10}$).
    The same apparatus is used for single-ion/single-photon experiments.
  }
  \label{F:Entangler}
\end{figure}

The projective measurement of the photons is performed with a partial Bell state analyser, consisting of a 50:50 non-polarising beam splitter (NPBS) and polarising beam splitters (PBSs) on each output arm.
All four output channels are monitored by avalanche photodiodes (APDs, quantum efficiency 65\%~\footnote{Laser Components COUNT BLUE, quantum efficiency measured at \SI{422}{\nano\metre}.}), as shown in Fig.~\ref{F:Entangler}.
Spatial mode matching of the photons from each system at the NPBS is aided by recoupling the light into AR-coated single-mode optical fibres before the APDs (coupling efficiency $\approx 90\%$).
The total click efficiency into all APDs is typically 2.1\%\ and 2.4\%\ for Alice and Bob, respectively. These measured efficiencies imply that the mode matching into the first fibre is $\gtrsim 50\%$ of the theoretical optimum [8.0\%, Fig.~\ref{F:Collection}(b)].

We first characterise the entanglement between the ion and emitted photon for each of the trap systems, using one detector in the apparatus shown in Fig.~\ref{F:Entangler}.
We perform full tomography of the combined ion-photon state by independently rotating each qubit.
Rotations of the ion state are performed on the optical qubit using the \SI{674}{\nano\metre} laser after mapping $\up$ to $\dstate$ with a $\pi$ pulse.
Rotations of the photon state are performed using the wave plates on the Bell state analyser.
An overcomplete set of ion and photon measurements is used to characterise the entangled ion-photon state, and to calculate the maximum-likelihood estimate (MLE) of the composite density matrix.
The density matrices obtained indicate a fidelity of \alicefidelity{} [\bobfidelity{}] with the maximally entangled state, at an average rate of \alicerate{} (\bobrate{}) for the Alice (Bob) system.

Ion qubit rotation errors account for $\approx 0.6\%$ of the total error, at $\approx 0.3\%$ per rotation.
We measure correlations of ion state with photon polarisation of $P\left(\uparrow \mid \! V \right) \approx P\left(\downarrow \mid \! H \right) \approx \num{0.995}$,
which includes the error from one $\pi$ pulse on the ion qubit.
This bounds the error due to all polarisation mixing effects to $\lesssim 0.2\%$.
Excited state preparation errors (preparing $\ket{\fslev{P}{1/2}{-1/2}}$ instead of $\ket{\fslev{P}{1/2}{+1/2}}$) depend on the polarisation impurity of both the optical pumping and pulsed excitation beams and are therefore suppressed.
The remaining 1.4\%\ error is attributed to ion qubit dephasing during the \SI{60}{\micro\second} delay between photon detection and tomography, and is expected to be due to noise in the applied magnetic field.

\begin{figure*}
  \includegraphics[width=\twocolfig]{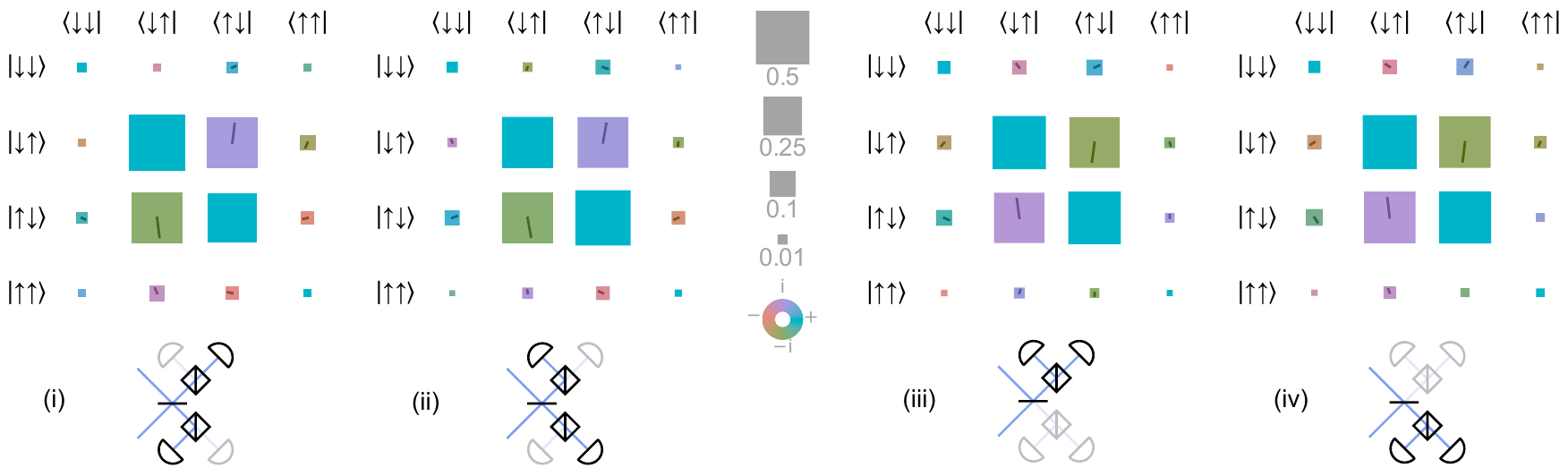}
  \caption{%
    The remote ion-ion density matrices and corresponding herald patterns (i)--(iv) as per the detector arrangement in Fig.~\ref{F:Entangler}.
    Detector clicks on opposing sides of the 50:50 beamsplitter (i), (ii) herald projection into $\ket{\Psi^{-}_{\rm{ion}}}$, while clicks on the same side (iii), (iv) herald $\ket{\Psi^{+}_{\rm{ion}}}$.
    The average fidelity of all four patterns to the nearest maximally entangled state is \abfidelity{}, at a heralded rate of \abrate{}.
    (In this diagram the area of each square gives the magnitude of the matrix element, with the color representing the complex phase, according to the key shown.
    A ``clock hand'' also indicates the phase on the same color wheel.
    See the Supplemental Material for numerical information.)
  }
  \label{F:Matrices}
\end{figure*}

To entangle the two remote ion qubits, we erase the path information of photons entangled with each ion and subsequently project the ion-ion state via a destructive measurement on the photon polarisations.
A coincidence detection on an appropriate pair of detectors heralds one of two Bell states: $\ketfixed{\Psi^{+}_{\rm{photon}}} := \left(\ket{VH} + \ket{HV}\right) / \sqrt{2}$ if the detectors are on the same output port of the NPBS, and $\ketfixed{\Psi^{-}_{\rm{photon}}} := \left(\ket{VH} - \ket{HV}\right) / \sqrt{2}$ if the detectors are on different output ports.
Detection of $\ketfixed{\Psi^\pm_{\rm{photon}}}$ projects the ions correspondingly into $\ketfixed{\Psi^\pm_{\rm{ion}}} := \left(\ket{\uparrow\downarrow} \pm e^{i \phi} \ket{\downarrow\uparrow}\right) / \sqrt{2}$, where the phase $\phi$ is stable~\cite{Stephenson2019} and can be transformed to zero with local operations.

The probability of successfully heralding an entanglement event is given~\cite{Hucul2015} by
\begin{equation}
  P = p_{\rm{Bell}} \left[ P_{\downarrow} P_e P_S \bar{P}_{\rm{click}} \right]^2,
  \label{E:Probability}
\end{equation}
where $p_{\rm Bell} = 1/2$ because we have valid heralds only for two of the four possible two-photon Bell states, $P_{\downarrow} \approx \num{0.99}$ is the probability of preparing the correct ground state before excitation, $P_e \approx \num{0.97}$ is the population transferred to the excited state by the pulsed excitation beam, $P_S \approx \num{0.95}$ is the probability of decaying to the $\lev{S}{1/2}$ ground states and $\bar{P}_{\rm{click}} \approx 0.023$ is the average probability of detecting a photon emitted by an ion.
We measure $P = \absuccessp{}$; given the average attempt rate (including Doppler cooling) of \SI{833}{\kilo\hertz}, $P$ yields a heralded ion-ion entanglement rate of \abrate{}.

After detection of a two-photon herald, we perform two-qubit tomography to verify the entangled state, using a series of single-qubit rotations and projective measurements~\footnote{See Supplemental Material for a more detailed discussion of methods and data analysis.}.
The MLE ion-ion state is calculated for each of the four herald patterns individually, as shown in Fig.~\ref{F:Matrices}, indicating an average fidelity of \abfidelity{} to the closest maximally entangled state~\cite{Badziag2000}.

The total ion-ion infidelity is dominated by errors in the ion-photon fidelity from each trap as described above, totaling 4.4\%, which includes errors in the ion-qubit rotations, ion dephasing, and polarisation mixing effects.
Additional infidelities include: the measured imperfections of the beamsplitters in the Bell state analyser, $\lesssim 0.17\%$; temporal misalignment of the photons, $\lesssim 0.13\%$; and dark counts, which contribute $\lesssim 0.05\%$ despite a relatively high dark count rate of \ish\SI{60}{\per\second} per APD.
The error due to mismatch of the photon modes at the NPBS is bounded by the measured fidelity to $< 1.3(5)\%$, which is approximately consistent with independent measurements.

In summary, we have used a new combination of collection geometry and excitation scheme to demonstrate remote entanglement between two atomic ion qubits at much higher rates and fidelities than previously measured.
The dominant infidelities arose from single-ion manipulations and spin decoherence, due to noise in the applied magnetic field and other known technical issues.
An order of magnitude of rate improvement is feasible by reducing latencies and the duration of state preparation shown in Fig.~\ref{F:SrLevels}.
Further rate gains could exploit the quadratic dependence on detection efficiency $P_{\rm{click}}$ indicated by Eq.~(\ref{E:Probability}), by using detectors of higher quantum efficiency, improving the mode matching into the fibres or using higher numerical aperture lenses to increase collection efficiency.
Significantly greater increases could in principle be realised by the use of a mirror close to the ion~\cite{Shu2010,Fischer2014}, or via the Purcell enhancement provided by an optical cavity~\cite{Cirac1997,Kim2011,Stute2012,Stute2013}.
Typical optical fibre losses at \SI{422}{\nano\metre} are \SI{30}{\decibel\per\kilo\metre}; frequency down conversion to the telecommunications \emph{C} band (\SI{1550}{\nano\metre})~\cite{Wright2018} would allow the distribution of entanglement over much larger distances than in this experiment.
The measured structure of the remote state produced is such that only two entangled pairs would be needed to distill a single remote entangled pair at or above 99\%\ fidelity~\footnote{S.\ Benjamin (private communication).}.
This would allow the photonic link to approach the performance of state-of-the-art local operations, enabling a variety of quantum networking applications.

We would like to thank Peter Maunz (Sandia National Laboratories) for supplying HOA2 ion traps, and the developers of ARTIQ~\cite{ARTIQ}.
B.\ C.\ N.\ and K.\ T.\ acknowledge funding from the U.K.\ National Physical Laboratory and the Defence Science and Technology Laboratory respectively.
C.\ J.\ B.\ acknowledges support from a UKRI FL Fellowship, and is a Director of Oxford Ionics Ltd.
This work was supported by the U.K.\ EPSRC ``Networked Quantum Information Technology'' Hub and the E.U.\ Quantum Technology Flagship Project AQTION (No.\ 820495).

{}

\bibliography{library}
\end{document}


\title{Supplemental material for `High-rate, high-fidelity entanglement of qubits across an elementary quantum network'}

\author{L.~J.~Stephenson}
\thanks{These two authors contributed equally}
\author{D.~P.~Nadlinger}
\thanks{These two authors contributed equally}
\author{B.~C.~Nichol}
\author{S.~An}
\author{P.~Drmota}
\author{T.~G.~Ballance}
\thanks{Present address: ColdQuanta UK Ltd, Oxford}
\author{K.~Thirumalai}
\author{J.~F.~Goodwin}
\author{D.~M.~Lucas}
\author{C.~J.~Ballance}
\email{chris.ballance@physics.ox.ac.uk}
\affiliation{Department of Physics, University of Oxford, Clarendon Laboratory, Parks Road, Oxford OX1 3PU, U.K.}

\date{May 12, 2020}

\maketitle

\section{S1. Photon collection}

To collect \SI{422}{\nm} photons emitted by the two ions, custom-designed lens objectives manufactured by Photon Gear \footnote{Photon Gear Inc., Ontario, NY, United States} are used. They provide near-diffraction-limited performance at an input-side numerical aperture of $0.6$ (specified: $\leq 0.08$ waves rms at \SI{403.1}{nm}). The image-side numerical aperture is $0.09$, designed to match the ion emission into standard, commercially available step-index fused-silica fibres.

The objective lens is mounted on a five-axis stage for fine control over translational and rotational alignment. After fitting the observed point spread function with a low-order Zernike polynomial model similar to ref.~\cite{Wong-Campos2016}, a cylindrical lens is inserted near the image plane to correct for residual aberrations likely caused by varying thickness across the vacuum viewport.

The tip of the collection fibre is mounted on a three-axis positioning stage, which can be adjusted using open-loop piezo actuators to correct for mechanical relaxation and slow, thermal drifts. Automatically tracking the optimum in observed fluorescence count rate allows us to retain good fibre-coupling efficiency over many days. (The short-term passive stability of the system was sufficient for typical experiments, such as the for the data presented in the main text, to be conducted without interruptions for realignment.)

\section{S2. Ion-photon tomography}

In the following, we discuss the experiments performed to separately characterise the ion-photon interface of each ion trap node, that is, the quantum state tomography measurements of the joint ion-photon state following picosecond laser excitation.

To implement the necessary measurements on the photonic state, we use the same Bell state analyser apparatus as for the remote entanglement experiment (see Fig.~3 in the main manuscript), but allow only light from one trap to enter the system. We determine the fast axes and retardances of the individual waveplates separately, and then run the experiment while varying their orientation in a motorised mount, reading out the ion state after a click has been observed on APD0.

We choose to analyse the ion state for all combinations of fast-axis angles of $\{ 0, \pi/4 \}$ for the quarter-wave plate, and $\{ 0, \pi/8, \pi/4, 3\pi/8 \}$ for the half-wave plate, defined with respect to the measurement polariser. This implements an over-complete set of projectors for measuring the photonic state. (For ideal waveplates, several of these choices would give rise to the same measurement projector.)

To analyse the ion state, we execute a sequence of \SI{674}{\nm} laser pulses to map part of the population into $\ket{\fslev{D}{5/2}{-3/2}}$ and optionally apply extra $\pi/2$-rotations around an axis in the $xy$ plane. As the phase of the ion-photon state in the laboratory frame is set with the detection of the photon, these analysis sequences are triggered at a fixed time offset from the avalanche photodetector click (with $\ish \SI{1}{\ns}$ precision).

\begin{figure*}[t]
	\centering
	\includegraphics[trim=1cm 2cm 1cm 0, width=\textwidth]{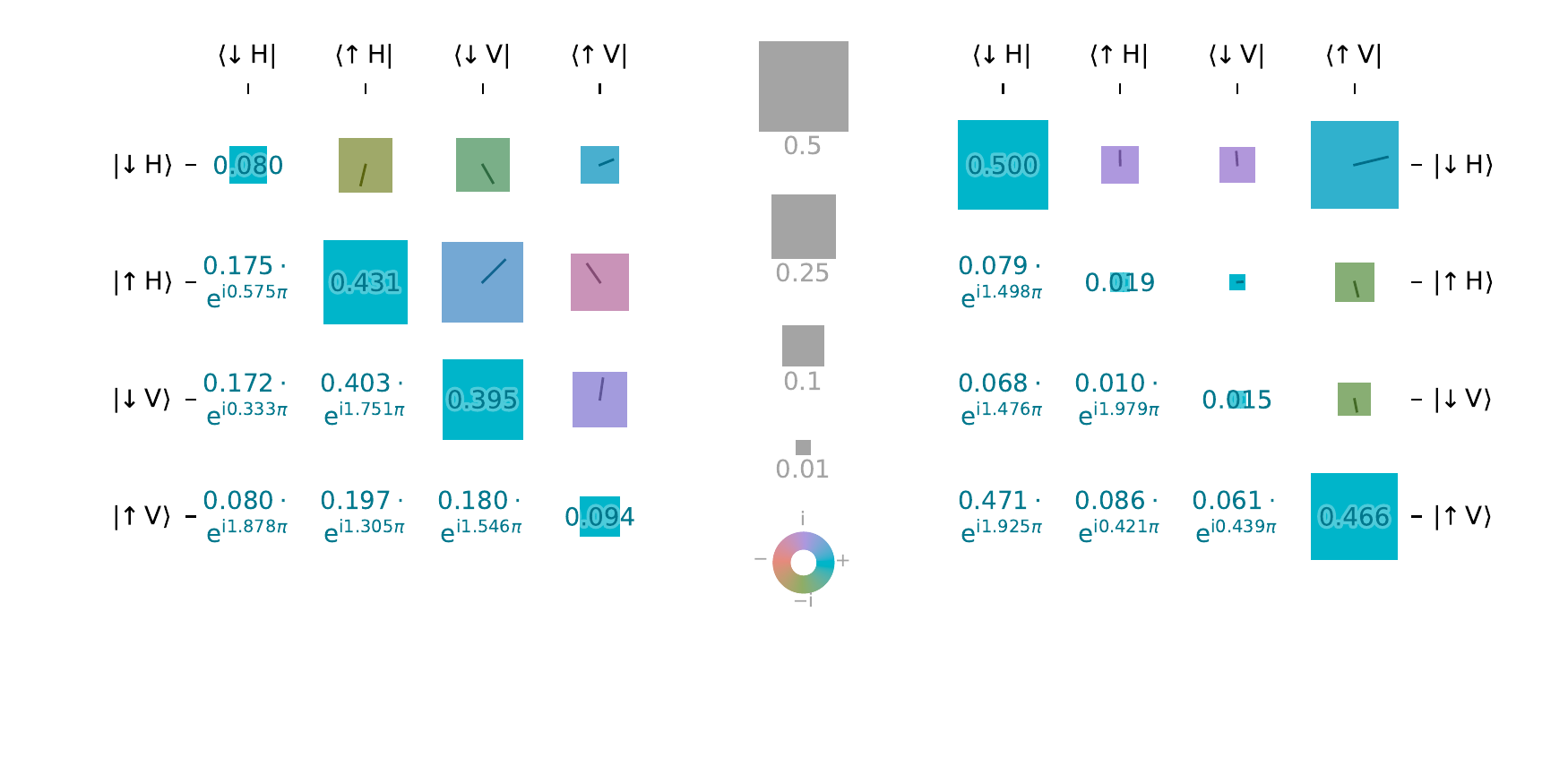}
	\caption{Maximum-likelihood estimates of the joint ion-photon state produced in nodes Alice (\emph{left}) and Bob (\emph{right}), obtained from \num{624000} and \num{572000} individual state tomography measurements, respectively. The fully entangled fractions are $\mathcal{F}_{\mathrm{Alice}} = \alicefidelity{}$ and $\mathcal{F}_{\mathrm{Bob}} = \bobfidelity{}$, respectively, with the error given as the s.e.m.~from parametric bootstrapping.}
	\label{fig:ion-photon-unrotated}
\end{figure*}
\begin{figure*}[t]
	\centering
	\includegraphics[trim=1cm 2cm 1cm 0, width=\textwidth]{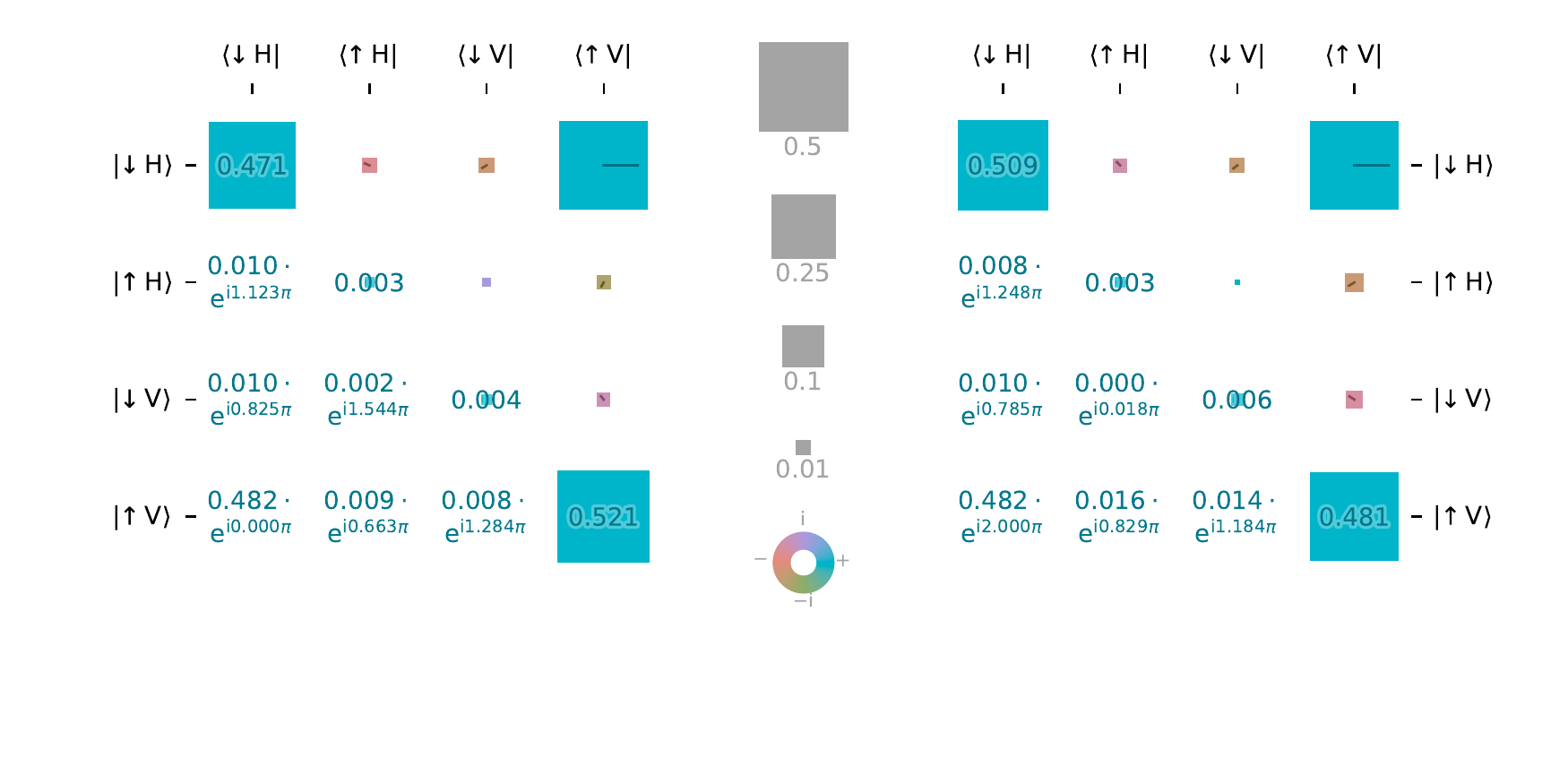}
	\caption{Maximum-likelihood ion-photon state estimates in nodes Alice (\emph{left}) and Bob (\emph{right}) from Fig.~\ref{fig:ion-photon-unrotated}, transformed post-hoc with unitary operations acting locally on the two subsystems to maximise the overlap with the Bell state $\ket{\Phi^+} := \left(\ket{\downarrow\mathrm{H}} + \ket{\uparrow\mathrm{V}}\right) / \sqrt{2}$.}
	\label{fig:ion-photon-rotated}
\end{figure*}

The results from two such tomography runs for Alice and Bob are shown in Fig.~\ref{fig:ion-photon-unrotated}. In total, \num{624000} (\num{572000}) copies of the ion-photon state were measured in Alice (Bob) for these datasets, distributed over \SI{27}{\min} (\SI{18}{\min}) of wall clock time. The density matrix estimates are obtained by direct numerical optimisation of the likelihood function over the $15 + 1$ real degrees of freedom parameterising the state and the overall collection efficiency. From those estimates, the fully entangled fraction \cite{Badziag2000}, that is, the fidelity to the nearest maximally entangled state, is computed as \alicefidelity{} and \bobfidelity{}, respectively, where the errors are given as the s.e.m.~obtained from parametric bootstrapping.

We make no attempt to control the polarisation rotation introduced by the single-mode fibre linking the ion traps with the polarisation state analyser setup. This results in an extra, a priori unknown, unitary transformation on the photonic part of the observed state. An arbitrary choice of phase reference is also made for the ion's state. To aid visual interpretation, the same MLE density matrix results are hence shown in Fig.~\ref{fig:ion-photon-rotated} after post-processing with local rotations to maximise their overlap with $\ket{\Phi^+} := \left(\ket{\downarrow\mathrm{H}} + \ket{\uparrow\mathrm{V}}\right) / \sqrt{2}$. Drifts in birefringence due to ambient temperature and mechanical stress lead to slow changes in the unitary induced by the fibre, but we typically observe it to be stable to $\gtish 99\%$ in state overlap over many days.

Imperfections in the \SI{674}{\nm} laser pulses and ion qubit decoherence due to magnetic field noise are expected to have a significant effect on the observed ion-photon correlations (see main text). While these effects can be characterised separately, we do not adjust the measurement projectors assumed in the tomography analysis for any such errors (nor for imperfect state discrimination). As such, the tomographic estimate provides only a lower bound for the isolated performance of the ion-photon interface.

\section{S3. Ion-ion tomography}

\begin{figure}[t]
	\centering
	\includegraphics{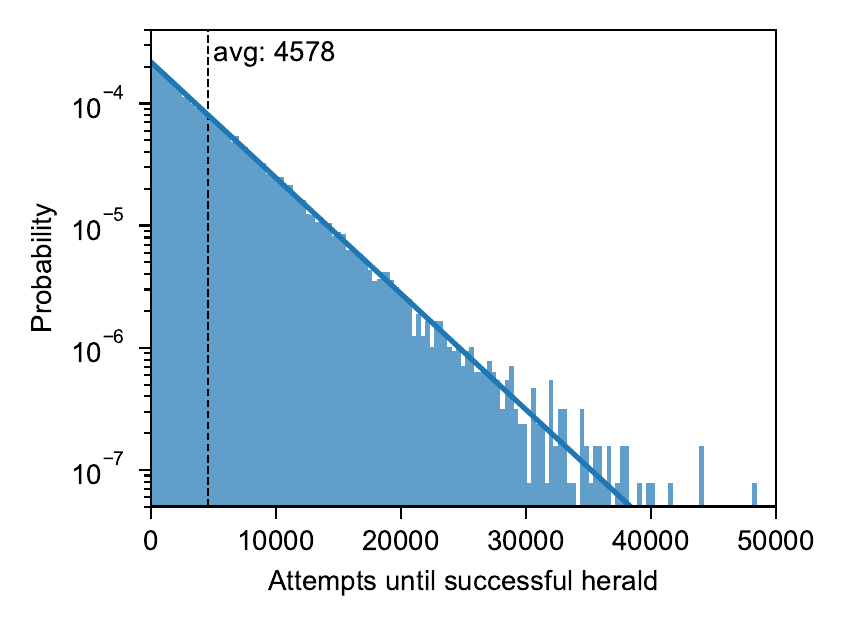}
	\caption{Histogram of the number of excitation attempts necessary until a two-photon herald event %
	was observed during acquisition of the ion-ion state tomography data presented in the main text. The solid line shows the exponential distribution corresponding to a Poisson process with success probability $1 / 4578 = \num{2.18e-4}$. The effective attempt rate is \SI{833}{kHz} after including the amortised cost of periodic cooling operations, resulting in an average remote entanglement generation rate of \abrate{}.}
	\label{fig:attempt-count}
\end{figure}

In the main text, we present a performance evaluation of the photonic entanglement generation procedure using tomography of the joint ion-ion quantum state.

To obtain tomography data, we first set the orientation of the waveplate pairs on the input to the Bell state analyser apparatus to compensate the birefringence of the fibre-optical links. We do this ahead-of-time in separate experiments by iteratively optimising the waveplate orientations for each node to maximise the probability of obtaining the ion $\down$ state after detecting a photon on APD0. This way, the polarisations incident on the non-polarising beam splitter for either decay channel are matched. Choosing equal Zeeman qubit frequencies by equalising the magnetic field strengths then removes any dependence of the resulting state on the photon detection times.

The state of each ion qubit is measured along the $x$-, $y$- or $z$-axis of the Bloch sphere, giving rise to nine possible measurement settings, each corresponding to a projective measurement onto the four eigenvectors of a tensor product of Pauli matrices. For the data presented in the main text, each measurement is performed on \num{1000} heralded states. For each attempt, we permit all four heralding detector combinations and record which occurred, to later separate them for data analysis \footnote{This separation is merely to diagnose unexpected differences between the herald patterns; in an application making use of the remotely entangled state, the click pattern dependence would immediately be corrected by inserting additional local phase rotations into the computation as necessary.}. We iterate four times through a random permutation of basis combinations, analysing a total of \num{36000} ion-ion states.

In total, the data acquisition time was approximately \SI{10}{min}. A total of \num{165321248} entanglement attempts were made, with the individual number of trials until success well-described by an exponential distribution, shown in Fig.~\ref{fig:attempt-count}.

\begin{figure*}[t]
    \centering
    \includegraphics{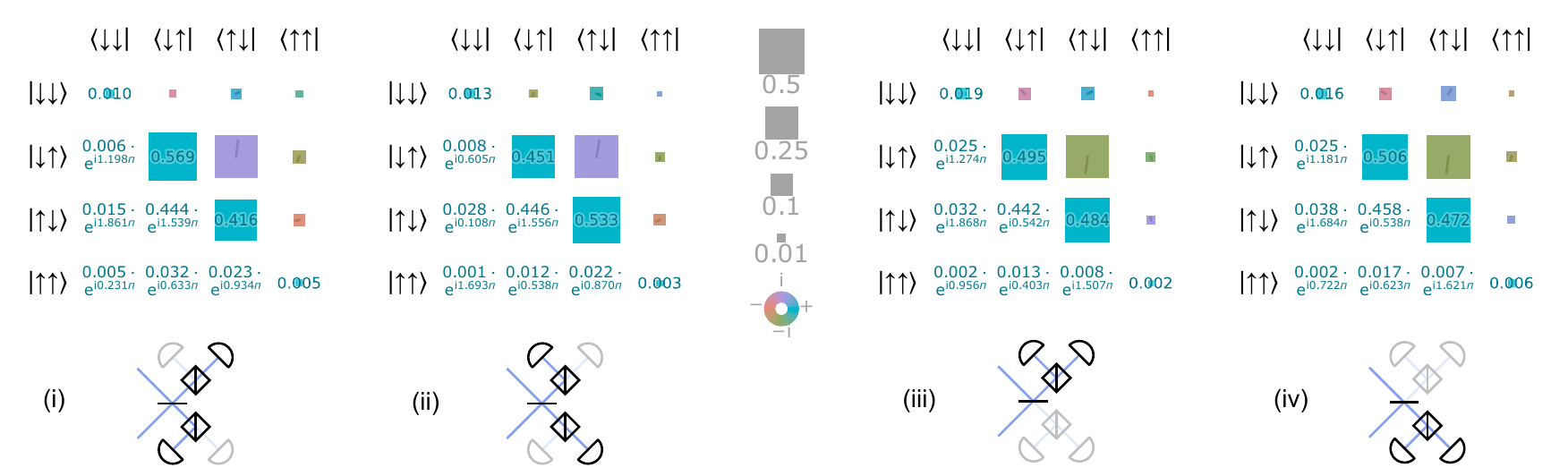}
    \caption{Maximum-likelihood ion-ion state estimates heralded by the four different photon detector click patterns. (This is a duplicate of Fig.~4 in the main text, but with the magnitude and phase of the matrix elements shown numerically.)}
    \label{fig:ion-ion-matrices}
\end{figure*}

\begin{figure*}[t]
	\centering
	\includegraphics{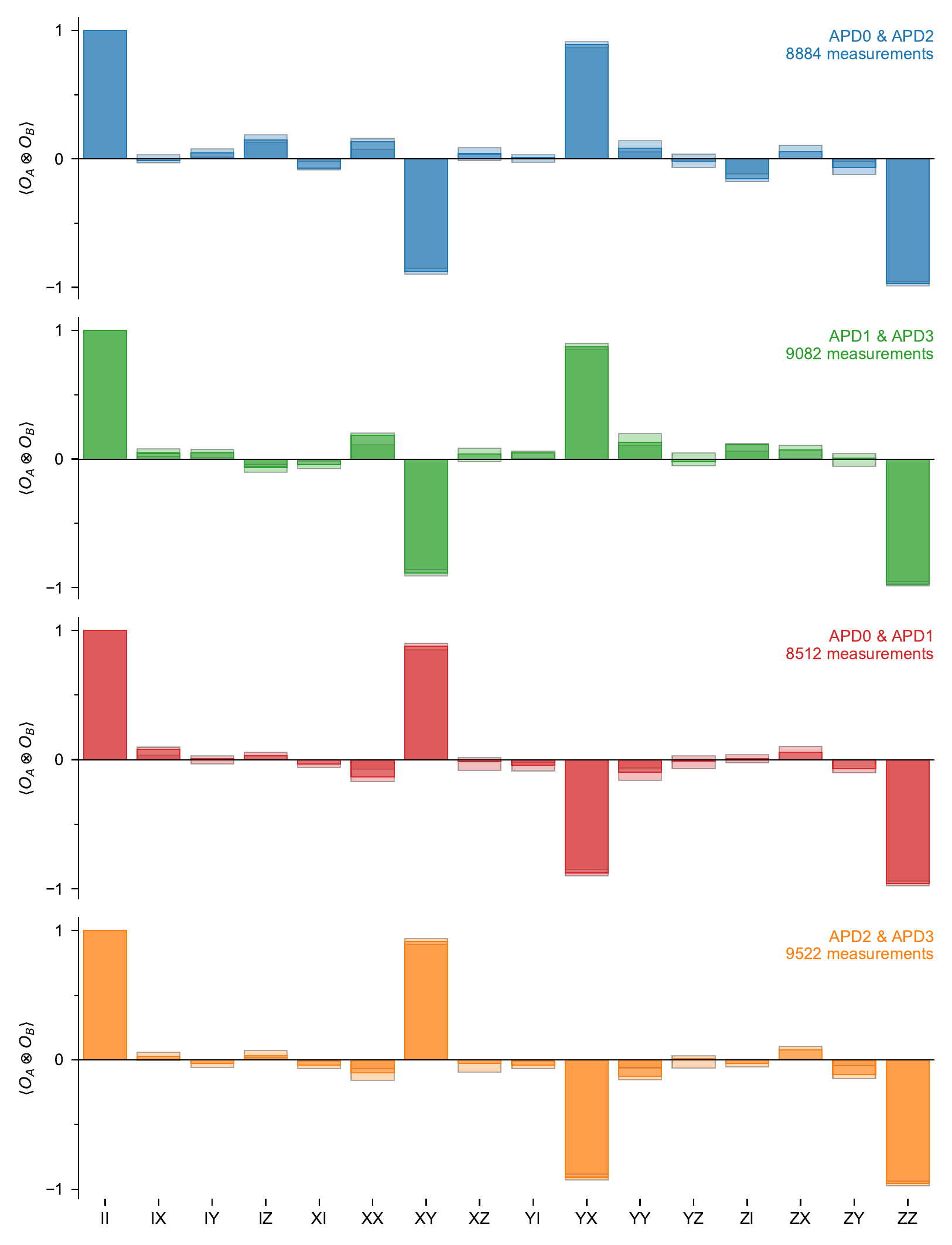}
	\caption{Ion-ion tomography data for the states heralded by the four different photon detector click patterns. For each two-qubit Pauli operator, the solid bars show the experimental measurements (averages directly computed from the raw data from Table~\ref{tab:ion-ion-counts}), whereas transparent bars denote the 95\%\ confidence interval of the respective expected value from the maximum-likelihood tomography estimate (obtained via bootstrapping).}
	\label{fig:ion-ion-data}
\end{figure*}

To obtain an estimate for the quantum state of the system conditioned on each of the four detector click patterns, we perform maximum-likelihood state tomography using a diluted fixed point iteration algorithm \cite{Rehacek2007}, which we also verify using direct numerical maximisation of the likelihood function. The resulting density matrices are in good agreement with the observed correlations; a comparison of observations and model is shown in Fig.~\ref{fig:ion-ion-data}. (There is a slight imbalance of observation counts between the four click patterns due to small differences in detector efficiencies.)

The quoted figures of merit (fidelity $\mathcal{F} = 0.940(5)$, entanglement of formation $E_F = 0.838(16)$) are computed from the density matrices and averaged over the detector click patterns. To estimate the statistical error, both parametric bootstrapping (simulated experiments from the MLE density matrices) and non-parametric bootstrapping (resampling of the experimental outcomes) are used, and found to give indistinguishable results; quoted as the s.e.m.~above and in the main text. The size of the confidence intervals was also verified against a Bayesian model \cite{Faist2016}, in which a Hilbert--Schmidt uniform prior for the density matrix is assumed and the posterior distribution for each figures of merit is sampled using a Markov-chain Monte-Carlo algorithm (Metropolis--Hastings, using the implementation from ref.~\cite{tomographer}).

No attempt was made to correct for the known qubit manipulation and state readout imperfections, which are a consequence of technical noise as discussed in the main text. As such, the quoted fidelities describe the whole-system performance and should be regarded as an upper bound for the link error.

\bibliography{library}

\begin{table*}
    \subfloat[Number of observations per Pauli eigenstate for detectors APD0 and APD2, for a total of \num{8884} clicks.]{%
        \begin{minipage}{0.45\textwidth}
            \centering
            \begin{tabular}{cc|cccc}
                $O_A$ & $O_B$ & $+_A+_B$ & $-_A+_B$ & $+_A-_B$ & $-_A-_B$ \\
                \hline
                Z & Z & 4 & 424 & 564 & 10 \\
                Z & X & 255 & 217 & 297 & 194 \\
                Z & Y & 283 & 218 & 316 & 193 \\
                X & Z & 212 & 222 & 271 & 263 \\
                X & X & 259 & 236 & 227 & 289 \\
                X & Y & 28 & 463 & 477 & 26 \\
                Y & Z & 178 & 213 & 288 & 285 \\
                Y & X & 424 & 32 & 32 & 524 \\
                Y & Y & 249 & 223 & 193 & 295 \\
            \end{tabular}
        \end{minipage}
    }
    \hspace{0.02\textwidth}
    \subfloat[Number of observations per Pauli eigenstate for detectors APD1 and APD3, for a total of \num{9082} clicks.]{%
        \begin{minipage}{0.45\textwidth}
            \centering
            \begin{tabular}{cc|cccc}
                $O_A$ & $O_B$ & $+_A+_B$ & $-_A+_B$ & $+_A-_B$ & $-_A-_B$ \\
                \hline
                Z & Z & 3 & 533 & 425 & 11 \\
                Z & X & 210 & 256 & 233 & 287 \\
                Z & Y & 204 & 294 & 232 & 252 \\
                X & Z & 263 & 278 & 235 & 231 \\
                X & X & 273 & 239 & 228 & 332 \\
                X & Y & 38 & 508 & 425 & 26 \\
                Y & Z & 254 & 280 & 267 & 250 \\
                Y & X & 451 & 30 & 27 & 495 \\
                Y & Y & 287 & 233 & 180 & 312 \\
            \end{tabular}
        \end{minipage}
    }

    \subfloat[Number of observations per Pauli eigenstate for detectors APD0 and APD1, for a total of \num{8512} clicks.]{%
        \begin{minipage}{0.45\textwidth}
            \centering
            \begin{tabular}{cc|cccc}
                $O_A$ & $O_B$ & $+_A+_B$ & $-_A+_B$ & $+_A-_B$ & $-_A-_B$ \\
                \hline
                Z & Z & 1 & 449 & 459 & 18 \\
                Z & X & 206 & 259 & 249 & 236 \\
                Z & Y & 247 & 260 & 224 & 185 \\
                X & Z & 234 & 237 & 260 & 251 \\
                X & X & 224 & 250 & 257 & 192 \\
                X & Y & 467 & 28 & 30 & 399 \\
                Y & Z & 221 & 249 & 220 & 263 \\
                Y & X & 21 & 441 & 450 & 36 \\
                Y & Y & 227 & 301 & 259 & 202 \\
            \end{tabular}
        \end{minipage}
    }
    \hspace{0.02\textwidth}
    \subfloat[Number of observations per Pauli eigenstate for detectors APD2 and APD3, for a total of \num{9522} clicks.]{%
        \begin{minipage}{0.45\textwidth}
            \centering
            \begin{tabular}{cc|cccc}
                $O_A$ & $O_B$ & $+_A+_B$ & $-_A+_B$ & $+_A-_B$ & $-_A-_B$ \\
                \hline
                Z & Z & 5 & 531 & 544 & 19 \\
                Z & X & 270 & 321 & 292 & 218 \\
                Z & Y & 271 & 278 & 309 & 234 \\
                X & Z & 262 & 236 & 282 & 263 \\
                X & X & 221 & 264 & 295 & 214 \\
                X & Y & 535 & 21 & 29 & 500 \\
                Y & Z & 240 & 261 & 242 & 289 \\
                Y & X & 20 & 511 & 482 & 24 \\
                Y & Y & 240 & 305 & 267 & 227 \\
            \end{tabular}
        \end{minipage}
    }
    \caption{Number of observations of each eigenstate for all tensor-product combinations $O_A \otimes O_B$ of Pauli operators in the ion-ion tomography experiment, for each of the four heralding detector patterns. This constitutes the complete input to the maximum-likelihood estimation procedure used to derive the density matrices presented the main text and Fig.~\ref{fig:ion-ion-matrices} (and, by extension, the quoted fidelities).}
    \label{tab:ion-ion-counts}
\end{table*}%